\input harvmac.tex
\vskip 1.5in
\Title{\vbox{\baselineskip12pt
\hbox to \hsize{\hfill}
\hbox to \hsize{\hfill WITS-CTP-053}}}
{\vbox{
	\centerline{\hbox{Gravitational Couplings of Higher Spins
		}}\vskip 5pt
        \centerline{\hbox{from String Theory
		}} } }
\centerline{Dimitri Polyakov\footnote{$^\dagger$}
{dimitri.polyakov@wits.ac.za}}
\medskip
\centerline{\it National Institute for Theoretical Physics (NITHeP)}
\centerline{\it  and School of Physics}
\centerline{\it University of the Witwatersrand}
\centerline{\it WITS 2050 Johannesburg, South Africa}
\vskip .3in

\centerline {\bf Abstract}
We  calculate the interaction 3-vertex of two massless 
spin 3 particles with a graviton using vertex operators for
spin 3 fields in open string theory, constructed in our previous work.
The massless spin 3 fields are shown to interact
with the graviton through the linearized Weyl tensor, reproducing
the result by Boulanger, Leclercq and Sundell.
This is consistent with the general structure
of the non-Abelian 
$2-s-s$ couplings, implying that the minimal number of space-time
derivatives in the interaction vertices of
two spin s  and one spin 2 
particle is equal to $2s-2$.
\Date{May 2010}
\vfill\eject

\lref\bianchi{ M. Bianchi, V. Didenko, arXiv:hep-th/0502220}
\lref\sagnottia{A. Sagnotti, E. Sezgin, P. Sundell, hep-th/0501156}
\lref\sorokin{D. Sorokin, AIP Conf. Proc. 767, 172 (2005)}
\lref\fronsdal{C. Fronsdal, Phys. Rev. D18 (1978) 3624}
\lref\coleman{ S. Coleman, J. Mandula, Phys. Rev. 159 (1967) 1251}
\lref\haag{R. Haag, J. Lopuszanski, M. Sohnius, Nucl. Phys B88 (1975)
257}
\lref\weinberg{ S. Weinberg, Phys. Rev. 133(1964) B1049}
\lref\fradkin{E. Fradkin, M. Vasiliev, Phys. Lett. B189 (1987) 89}
\lref\skvortsov{E. Skvortsov, M. Vasiliev, Nucl.Phys.B756:117-147 (2006)}
\lref\skvortsovb{E. Skvortsov, J.Phys.A42:385401 (2009)}
\lref\mva{M. Vasiliev, Phys. Lett. B243 (1990) 378}
\lref\mvb{M. Vasiliev, Int. J. Mod. Phys. D5
(1996) 763}
\lref\mvc{M. Vasiliev, Phys. Lett. B567 (2003) 139}
\lref\brink{A. Bengtsson, I. Bengtsson, L. Brink, Nucl. Phys. B227
 (1983) 31}
\lref\deser{S. Deser, Z. Yang, Class. Quant. Grav 7 (1990) 1491}
\lref\bengt{ A. Bengtsson, I. Bengtsson, N. Linden,
Class. Quant. Grav. 4 (1987) 1333}
\lref\boulanger{ X. Bekaert, N. Boulanger, S. Cnockaert,
J. Math. Phys 46 (2005) 012303}
\lref\metsaev{ R. Metsaev, arXiv:0712.3526}
\lref\siegel{ W. Siegel, B. Zwiebach, Nucl. Phys. B282 (1987) 125}
\lref\siegelb{W. Siegel, Nucl. Phys. B 263 (1986) 93}
\lref\nicolai{ A. Neveu, H. Nicolai, P. West, Nucl. Phys. B264 (1986) 573}
\lref\damour{T. Damour, S. Deser, Ann. Poincare Phys. Theor. 47 (1987) 277}
\lref\sagnottib{D. Francia, A. Sagnotti, Phys. Lett. B53 (2002) 303}
\lref\sagnottic{D. Francia, A. Sagnotti, Class. Quant. Grav.
 20 (2003) S473}
\lref\sagnottid{D. Francia, J. Mourad, A. Sagnotti, Nucl. Phys. B773
(2007) 203}
\lref\labastidaa{ J. Labastida, Nucl. Phys. B322 (1989)}
\lref\labastidab{ J. Labastida, Phys. rev. Lett. 58 (1987) 632}
\lref\mvd{L. Brink, R.Metsaev, M. Vasiliev, Nucl. Phys. B 586 (2000)183}
\lref\klebanov{ I. Klebanov, A. M. Polyakov,
Phys.Lett.B550 (2002) 213-219}
\lref\mve{
X. Bekaert, S. Cnockaert, C. Iazeolla,
M.A. Vasiliev,  IHES-P-04-47, ULB-TH-04-26, ROM2F-04-29, 
FIAN-TD-17-04, Sep 2005 86pp.}
\lref\sagnottie{A. Campoleoni, D. Francia, J. Mourad, A.
 Sagnotti, Nucl. Phys. B815 (2009) 289-367}
\lref\sagnottif{
A. Campoleoni, D. Francia, J. Mourad, A.
 Sagnotti, arXiv:0904.4447}
\lref\sagnottig{D. Francia, A. Sagnotti, J.Phys.Conf.Ser.33:57 (2006)}
\lref\taronna{M. Taronna, arXiv:1005.3061}
\lref\desera{C. Aragone, S. Deser, Nuovo Cim. B57 (1980) 33-49}
\lref\berends{F. Berends, J. W. van Holten, P. van Nieuwenhuizen,
B. de Wit, Phys. Lett. B83 (1979) 188}
\lref\metsaevb{R. Metsaev, Nucl. Phys. B759 (2006) 147-201}
\lref\selfa{D. Polyakov, Int.J.Mod.Phys.A20:4001-4020,2005}
\lref\selfb{ D. Polyakov, arXiv:0905.4858}
\lref\selfc{D. Polyakov, arXiv:0906.3663}
\lref\selfd{D. Polyakov, Phys.Rev.D65:084041 (2002)}
\lref\mirian{A. Fotopoulos, M. Tsulaia, Phys.Rev.D76:025014,2007}
\lref\extraa{I. Buchbinder, V. Krykhtin,  arXiv:0707.2181}
\lref\extrab{I. Buchbinder, V. Krykhtin, Phys.Lett.B656:253-264,2007}
\lref\extrac{X. Bekaert, I. Buchbinder, A. Pashnev, M. Tsulaia,
Class.Quant.Grav. 21 (2004) S1457-1464}
\lref \extrad{I. Buchbinder, A. Pashnev, M. Tsulaia,
arXiv:hep-th/0109067}
\lref\extraf{I. Buchbinder, A. Pashnev, M. Tsulaia, 
Phys.Lett.B523:338-346,2001}
\lref\extrag{I. Buchbinder, E. Fradkin, S. Lyakhovich, V. Pershin,
Phys.Lett. B304 (1993) 239-248}
\lref\extrah{I. Buchbinder, A. Fotopoulos, A. Petkou, 
 Phys.Rev.D74:105018,2006}
\lref\bonellia{G. Bonelli, Nucl.Phys.B {669} (2003) 159}
\lref\bonellib{G. Bonelli, JHEP 0311 (2003) 028}
\lref\ouva{C. Aulakh, I. Koh, S. Ouvry, Phys. Lett. 173B (1986) 284}
\lref\ouvab{S. Ouvry, J. Stern, Phys. Lett.  177B (1986) 335}
\lref\ouvac{I. Koh, S. Ouvry, Phys. Lett.  179B (1986) 115 }
\lref\hsself{D. Polyakov, arXiv:0910.5338}
\lref\nba{N. Boulanger, S. Leclercq, JHEP 11 (2006) 034}
\lref\nbb{N. Boulanger, S. Leclercq, P. Sundell, JHEP 0808 (2008) 056}
\lref\henna{G. Barnich, M. Henneaux, Phys. Lett. B311 (1993) 123-129}
\lref\hennb{M. Henneau, Contemp. Math. 219 (1998) 93}
\lref\fvb{E.S. Fradkin, M. A. Vasiliev, Nucl. Phys. B291 (1987) 141}

\centerline{\bf  1. Introduction}

Higher spin field theory is a fascinating subject that has attracted
a lot of attention in recent years. Consistent description of
theories with interacting higher spins 
is well-known to be a difficult problem, in particular 
because such theories need to
involve powerful gauge symmetries, sufficient to remove the negative norm 
states. As a result, introducing higher spin interactions in a consistent way
is a complicated issue with many challenging obstacles, despite certain
progress in this direction over recent years
~{\fradkin, \mva, \mvb, \brink, \deser, \boulanger, \metsaev, \siegel, 
\nicolai, \damour, \sagnottib, \sagnottia, \mvd, \mve, \sagnottie, \sagnottif,
\sagnottig, \taronna} 
String theory, on the other hand, appears to be a particularly efficient 
tool to describe interacting higher spins as many complex issues
in higher spin field theories are settled very naturally in the 
string-theoretic approach.
In particular, in our recent work ~{\hsself} we have been able to construct
the emission vertices for the massless higher spin fields
with the integer spin values $3\leq{s}\leq{9}$.
It has been shown that BRST-invariance conditions for these operators
lead to standard Fierz-Pauli constraints on the space-time fields 
in the Fronsdal's approach ~{\fronsdal}; BRST-nontriviality conditions, in turn,
 entail the gauge transformations for the space-time fields. 
The interaction terms of the higher spin fields could then be obtained
from the scattering amplitudes on the string theory side; since
the gauge transformations of the space-time fields shift the vertex
operators by a trivial part (not contribuding to correlation functions),
the interaction terms, obtained this way, will be gauge-invariant by 
construction.

In this work we concentrate on analyzing the important case
of the spin 3 field's gravitational
interaction, using  the vertex operators constructed in our previous work.
Since $s=3$ vertices are the open string operators, the spin 3 - graviton
coupling is described by the appropriate disc amplitude in string theory, 
with two
$s=3$ operators being on the boundary and the graviton in the bulk of the
disc.

The amplitude, considered in this paper, is shown to reproduce some
important properties of interacting higher spins, known from the 
field-theoretic
approach ~{\nba, \nbb, \henna, \hennb, \fvb}

In particular, while the gravitational couplings 
of lower spin fields contain at most two space-time derivatives
in the Lagrangian, 
 the gravitational couplings of the higher spin
fields are known to have a very different behaviour
~{\desera, \berends, \metsaevb, \nbb}.
 For example,
there exists a unique nonabelian
 interaction 3-vertex of a graviton with 2 particles
of spin $s$  
 containing $2s-2$ space-time derivatives, related to the flat limit of the
Fradkin-Vasiliev vertex in AdS.
In a particularly important $s=3$ case the relation to the Fradkin-Vasiliev
vertex can be established explicitly ~{\nba, \nbb} and the interaction
of two $s=3$ particles with a graviton can be expressed in terms of 
linearized Weyl tensor ~{\nba, \nbb}.
The purpose of this paper is to obtain this interaction from the string
theory side, using the vertex operators constructed in our previous work.
Computing the disc amplitude of two $s=3$ vertex operator with a graviton,
we recover the structure of the gravitational coupling
of massless spin 3 particles established in  ~{\nba,\nbb}, 
including the appearance
of the Weyl tensor. The answer for the cubic interaction vertex 
that follows from the string theory computation, agrees with the result
of ~{\nba, \nbb}, modulo overall normalization coefficient
and partial gauge fixing.
The rest of the paper is organized as follows. 
In the Section 2 we recall the basic properties
of the higher spin vertex operators, including the homotopy
transformations needed to construct these operators at positive ghost pictures.
In section 3 we perform the computation of the disc amplitude, showing
the resulting interaction to agree with the coupling structure given in 
~{\nba, \nbb}.
In the concluding section we briefly discuss implications of our results,
outlining future calculations.
 
\centerline{\bf Computation of Spin 3 - Graviton Amplitude}

The vertex operator for a massless symmetric spin 3 particle is given by
~{\hsself}:
\eqn\grav{\eqalign{
V_{s=3}(p)=H_{a_1a_2a_3}(p)c{e}^{-3\phi}\partial{X^{a_1}}\partial{X^{a_2}}
\psi^{a_3}e^{i{\vec{p}}{\vec{X}}}}}
in the unintegrated   $b-c$ picture and at the minimal negative 
$\beta-\gamma$ picture $-3$. 
Here $H_{a_1a_2a_3}(p)$ is symmetric tensor satisfying the on-shell
conditions
\eqn\grav{\eqalign{p^{a_1}H_{a_1a_2a_3}(p)=0\cr
\eta^{a_1a_2}H_{a_1a_2a_3}(p)=0\cr
p^2H_{a_1a_2a_3}(p)=0}}
as a consequence of the BRST-invariance constraints on the operator (1),
while the gauge transformations
\eqn\grav{\eqalign{H_{a_1a_2a_3}(p)\rightarrow
{H_{a_1a_2a_3}(p)}+\partial_{(a_1}\Lambda_{a_2a_3)}}}
with traceless symmetric $\Lambda$ parameter shift the vertex operator
(1) by a trivial part, not contributing to correlators.

To compute  a three-point
amplitude of two spin 3 particles with a graviton one also needs
the positive $\beta-\gamma$ picture $+1$ versions of $V_{s=3}$
in order to satisfy the ghost number anomaly cancellations.
These operators exist at integrated $b-c$-picture only and can be obtained
by replacing $-3\phi\rightarrow\phi$ with the subsequent homotopy
transformation described in ~{\hsself}, in order to ensure its BRST-invariance.
As it has been shown ~{\hsself}, given the higher spin field
described by ghost-dependent vertex operator
at minimal negative superconformal ghost picture $-n-2$ ( $n\geq{1}$),
\eqn\lowen{V_{-n-2}=ce^{-(n+2)\phi}F_{{{n^2}\over2}+n+1}(X,\psi)}
where (suppressing the space-time indices)
 $F_{{{n^2}\over2}+n+1}(X,\psi)$ the is matter
primary field of conformal dimension ${{n^2}\over2}+n+1$. 
To construct
the positive picture version of vertex operator describing such a state,
one starts with constructing the  charge 
\eqn\lowen{\oint{V_n}\equiv
\oint{{dz}}e^{n\phi}F_{{{n^2}\over2}+n+1}(X,\psi)}
The BRST operator is given by
\eqn\lowen{Q_{brst}=Q_1+Q_2+Q_3}

where

\eqn\grav{\eqalign{
Q_1=\oint{{dz}\over{2i\pi}}\lbrace{cT-bc\partial{c}}\rbrace\cr
Q_2=-{1\over2}\oint{{dz}\over{2i\pi}}\gamma\psi_a\partial{X^a}\cr
Q_3=-{1\over4}\oint{{dz}\over{2i\pi}}b\gamma^2}}

where T is the full stress-energy tensor.
The operator (1) commutes with $Q_1$ since it is a worldsheet integral
of dimension 1 and $b-c$ ghost number zero but doesn't commute with
$Q_2$ and $Q_3$. To make it BRST-invariant, one has to
add the correction terms by using the following procedure 
~{\selfb, \selfc}.
We write
\eqn\grav{\eqalign{\lbrack{Q_{brst}},V_n(z)\rbrack=\partial{U}(z)+W_1(z)
+W_2(z)}}
and therefore
\eqn\lowen{\lbrack{Q_{brst}},
\oint{dz}V_n{\rbrack}=\oint{{dz}}(W_1(z)+W_2(z))}
where 
\eqn\grav{\eqalign{U(z)\equiv{cV_n(z)}\cr
\lbrack{Q_1,V_n}\rbrack=\partial{U}\cr
W_1=\lbrack{Q_2,V_n}\rbrack\cr
W_2=\lbrack{Q_3},V_n\rbrack}}
Introduce the dimension 0 $K$-operator:
\eqn\lowen{K(z)=-4c{e}^{2\chi-2\phi}(z)\equiv{\xi}\Gamma^{-1}(z)}
satisfying
\eqn\lowen{\lbrace{Q_{brst}},K\rbrace=1}
It is easy to check that this operator has a non-singular
operator product with $W_1$:
\eqn\lowen{K(z_1)W_1(z_2)\sim{(z_1-z_2)^{2n}}Y(z_2)+O((z_1-z_2)^{2n+1})}
where $Y$ is some operator of dimension $2n+1$.
Then the complete BRST-invariant operator
can be obtained from $\oint{dz}V_n(z)$
by the following transformation:

\eqn\grav{\eqalign{
\oint{dz}{V_n}(z){\rightarrow}A_n(w)=\oint{dz}V_n(z)+{{1}\over{(2n)!}}
\oint{dz}(z-w)^{2n}:K\partial^{2n}{(W_1+W_2)}:(z)
\cr
+{1\over{{(2n)!}}}\oint{{dz}}\partial_z^{2n+1}{\lbrack}
(z-w)^{2n}{K}(z)\rbrack{K}\lbrace{Q_{brst}},U\rbrace}}
where $w$ is some arbitrary point on the worldsheet.
It is then straightforward to check the invariance
of $A_n$ by using some partial integration along with
the relation (12) as well as the obvious identity
\eqn\lowen{\lbrace{Q_{brst}},W_1(z)+W_2(z)\rbrace=
-\partial(\lbrace{Q_{brst}},U(z)\rbrace)}
Although the invariant operators $A_n(w)$ depend on an
arbitrary point $w$ on the worldsheet, this dependence
is irrelevant in the correlators
 since all the $w$ derivatives  of $A_n$ are BRST exact -
the triviality of the derivatives ensures that
 there will be no $w$-dependence in any correlation
functions involving $A_n$.
Equivalently, 
the positive picture representations $A_n$ (14) for higher spin operators
can also be obtained from minimal negative picture representations
$V_{-n-2}$ by straightforward, but technically more
cumbersome procedure 
by using  the combination of the picture-changing and 
the $Z$-transformation (the analogue of the picture-changing
for the $b-c$-ghosts).

Namely, the $Z$-operator, transforming the $b-c$ pictures (in particular,
mapping integrated vertices to unintegrated)
 given by ~{\selfa}
\eqn\lowen{Z(w)=b\delta(T)(w)=
\oint{dz}(z-w)^3(bT+4c\partial\xi\xi{e^{-2\phi}}T^2)(z)}
where $T$ is the full stress-energy tensor in RNS theory.
The usual picture-changing operator,
transforming the $\beta-\gamma$ ghost pictures, is given by
 $\Gamma(w)=:\delta(\beta)G:(w)
=:e^\phi{G}:(w)$.
Introduce the  $integrated$ picture-changing operators
$R_n(w)$ according to
\eqn\grav{\eqalign{R_{n}(w)=Z(w):\Gamma^{n}:(w)}}
where $:\Gamma^n:$ is the $n$th power of the standard
picture-changing operator:
\eqn\grav{\eqalign{
:\Gamma^{n}:(w)=:e^{{n}\phi}\partial^{n-1}G...\partial{G}G:(w)\cr
\equiv:\partial^{n-1}\delta(\beta)...\partial\delta(\beta)\delta(\beta):}}
Then the positive picture representations for the higher
spin operators $A_n$ can be obtained from the negative ones $V_{-n-2}$ (1)
by the transformation:
\eqn\lowen{A_n(w)=(R_2)^{n+1}(w){V_{-n-2}}(w)}

Since  both $Z$ and $\Gamma$ are BRST-invariant and nontrivial,
the $A_n$-operators by construction 
satisfy the BRST-invariance and non-triviality
conditions identical to those satisfied by their negative picture
counterparts
$V_{-2n-2}$ and therefore lead to the same Pauli-Fierz on-shell
conditions and the gauge symmetries for the higher spin fields.
Applying the above procedure to the $s=3$ vertex operator (1) (corresponding
to the $n=2$ case) we obtain the following expression for 
the positive $+1$ picture operators for $s=3$ massless particles.

\eqn\lowen{A_{s=3}(w)=\oint{{dz}}(z-w)^2U(z){\equiv}A_0+A_1+A_2+A_3+A_4+A_5+A_6+
A_7+A_8}
where
\eqn\lowen{A_0(w)={1\over2}H_{a_1a_2a_3}(p)\oint{dz}(z-w)^2
P^{(2)}_{2\phi-2\chi-\sigma}
e^{\phi}\partial{X^{a_1}}\partial{X^{a_2}}
\psi^{a_3}{e^{i{\vec{p}}{\vec{X}}}}(z)}
and
\eqn\lowen{A_8(w)=-12H_{a_1a_2a_3}(p)\oint{dz}(z-w)^2
\partial{c}c\partial\xi\xi{e^{-\phi}}
\partial{X^{a_1}}\partial{X^{a_2}}
\psi^{a_3}\rbrace{e^{i{\vec{p}}{\vec{X}}}}(z)}
have ghost factors proportional to
$e^\phi$ and $\partial{c}c\partial\xi\xi{e^{-\phi}}$ respectively
and the rest of the terms carry ghost factor proportional to
$c\xi$:
\eqn\grav{\eqalign{A_1(w)=-2H_{a_1a_2a_3}(p)\oint{dz}(z-w)^2
c\xi({\vec{\psi}}\partial{\vec{X}})
\partial{X^{a_1}}\partial{X^{a_2}}
\psi^{a_3}
{e^{i{\vec{p}}{\vec{X}}}}(z)\cr
A_2(w)=
-H_{a_1a_2a_3}(p)\oint{dz}(z-w)^2c\xi
\partial{X^{a_1}}\partial{X^{a_2}}
\partial{X^{a_3}}P^{(1)}_{\phi-\chi}
{e^{i{\vec{p}}{\vec{X}}}}(z)\cr
A_3(w)=
-H_{a_1a_2a_3}(p)\oint{dz}(z-w)^2c\xi
\partial{X^{a_1}}\partial{X^{a_2}}\partial^2{X^{a_3}}
{e^{i{\vec{p}}{\vec{X}}}}(z)\cr
A_4(w)=
2H_{a_1a_2a_3}(p)\oint{dz}(z-w)^2
c\xi\partial\psi^{a_2}P^{(1)}_{\phi-\chi})
\psi^{a_3}
{e^{i{\vec{p}}{\vec{X}}}}(z)\cr
A_5(w)=
2H_{a_1a_2a_3}(p)\oint{dz}(z-w)^2
c\xi\partial^2\psi^{a_2}
\psi^{a_3}
{e^{i{\vec{p}}{\vec{X}}}}(z)\cr
A_5(w)=
-2H_{a_1a_2a_3}(p)\oint{dz}(z-w)^2c\xi
\partial{X^{a_1}}\partial{X^{a_2}}(\partial^2{X^{a_3}}+
\partial{X^{a_3}}P^{(1)}_{\phi-\chi})
{e^{i{\vec{p}}{\vec{X}}}}(z)\cr
A_6(w)=2iH_{a_1a_2a_3}(p)\oint{dz}(z-w)^2
c\xi({\vec{p}}{\vec{\psi}})P^{(1)}_{\phi-\chi}
\partial{X^{a_1}}\partial{X^{a_2}}
\psi^{a_3}
{e^{i{\vec{p}}{\vec{X}}}}(z)\cr
A_7(w)=
2iH_{a_1a_2a_3}(p)\oint{dz}(z-w)^2
c\xi({\vec{p}}\partial{\vec{\psi}})
\partial{X^{a_1}}\partial{X^{a_2}}
\psi^{a_3}
{e^{i{\vec{p}}{\vec{X}}}}(z)
}}

\centerline{\bf 3. Calculation of the Disc Amplitude}

To evaluate the three-point amplitude of two $s=3$ particles
with a graviton on the disc,
 it is convenient to take both of the $s=3$ vertices at positive $+1$-picture
(integrated over the disc boundary), while taking the graviton at 
the disc origin unintegrated and at the left and right $\beta-\gamma$
ghost pictures $-1$ and $-2$ respectively:
\eqn\lowen{
V_{s=2}=\gamma_{m_1m_2}(p)c\bar{c}e^{-\phi-2\bar{\phi}}
\psi^{m_1}{\bar\partial}X^{m_2}(0,0)}
Furthermore, it shall be convenient to make conformal mapping from the
disc to the upper half-plane $(z,{\bar{z}})\rightarrow{(w,{\bar{w}})}$
using $w=-i{{z-i}\over{z+i}}$ so the the graviton's location is mapped
to $w=i$. 
As the ghost number anomaly cancellation requires that the
the total $\phi$-ghost number of correlation functions
has to be equal to $-2$, $\chi$-ghost number equal to 1  and the $\sigma$-ghost
number equal to 3,
 the three-point correlator
$<A_{s=3}(w_1)A_{s=3}(w_2)A_{s=2}(i)>$ is contributed by 
the terms
\eqn\grav{\eqalign{<A_{s=3}(w_1)A_{s=3}(w_2)A_{s=2}(i)>=
\sum_{j=1}^7<A_j(w_1)A_0(w_2)A_{s=2}(i)>}}
of (21), (23), (24).
So we shall calculate these terms one by one, using (23).
The cubic interaction vertex is  determined by the structure
constants given by the on-shell limit of (25), implying
$(p_ip_j)={0}$ 
In this paper , we are interested in the contributions to
the correlator (25) up to the terms quartic in momentum.
More precisely, our main result is that all the terms that are zero 
or quadratic in momentum vanish (as it is immediately clear from the
structure of the operators (21), (23), (24) 
that there are no terms linear or cubic in $p$)
and the lowest order terms are thus quartic in momentum 
(and hence in space-time
derivatives). As for the quartic order terms, we shall show that they combine
into the cubic interaction vertex manifestly dependent on the linearized
Weyl tensor, in agreement with the results of ~{\nba, \nbb}.

To start with, it is convenient to point out the ghost factor,
common for all the terms in (25).
It is given by:
\eqn\grav{\eqalign{G_{gh}(z_1,z_2,z,{\bar{z}})=
<ce^\chi(w_1)P^{(2)}_{2\phi-2\chi-\sigma}e^\phi(w_2)
ce^{-\phi}(z){\bar{c}}e^{-2{\bar{\phi}}}({\bar{z}})>
\cr
={{(z-z_1)(z-z_2)({\bar{z}}-z_1)({\bar{z}}-z_2)^2}\over{(z-{\bar{z}})}}
\cr\times\lbrace{{2\over{(z_1-z_2)^2}}+{1\over{({\bar{z}}-z_2)^2}}
+{1\over{(z-z_2)({\bar{z}}-z_2)}}-{1\over{(z-z_2)(z_1-z_2)}}
-{3\over{(z_1-z_2)({\bar{z}}-z_2)}}}\rbrace}}
Since the correlation functions of the
spin 3 operators $A(w)\sim\int{dz}(z-w)^2U(z)$ do not depend
on $w$,it is convenient to choose $w_{1,2}=\pm{i}$ for the first and
for the second $s=3$ operators. 
We start from the contributions to (25), quartic in the momentum.
The calculation of the first contribution, using the expressions (23), 
(24) gives
\eqn\grav{\eqalign{
<A_1(i)A_0(-i)V_{s=2}(z,-{\bar{z}})>_4=
H_{a_1a_2a_3}(p_1)H_{b_1b_2b_3}(p_2)\gamma_{m_1m_2}(p_3)
\cr\times
\int_{-\infty}^\infty
dz_1\int_{-\infty}^\infty{dz_2}(z_1-i)^2(z_2+i)^2{G_{gh}(z_1,z_2,i,-i)}
\cr\times
\lbrace{-}{{\eta^{a_4b_2}(-\eta^{a_4b_3}\eta^{a_3m_2}+\eta^{a_3b_3}\eta^{a_4m_2})}
\over{(z_1-z_2)^2(z_1-z)(z_2-{\bar{z}})}}
p_2^{a_1}p_2^{a_2}p_1^{b_1}p_1^{m_1}
\cr\times
({1\over{z_1-z_2}}-{1\over{z_1-{\bar{z}}}})^2
({1\over{z_1-z_2}}+{1\over{z_2-{\bar{z}}}})
({1\over{z_1-{\bar{z}}}}-{1\over{z_2-{\bar{z}}}})
\cr
{-}{{\eta^{a_4m_1}(-\eta^{a_4b_3}\eta^{a_3m_2}+\eta^{a_3b_3}\eta^{a_4m_2})}
\over{(z_1-{\bar{z}})^2(z_1-z)(z_2-{\bar{z}})}}
p_2^{a_1}p_2^{a_2}p_1^{b_1}p_1^{b_2}({1\over{z_1-z_2}}-{1\over{z_1-{\bar{z}}}})^2
({1\over{z_1-z_2}}+{1\over{z_2-{\bar{z}}}})^2
\cr
{-}{{\eta^{b_2m_1}(-\eta^{a_4b_3}\eta^{a_3m_2}+\eta^{a_3b_3}\eta^{a_4m_2})}
\over{(z_2-{\bar{z}})^2(z_1-{\bar{z}})(z_2-{{z}})}}
p_2^{a_1}p_2^{a_2}p_2^{a_4}p_1^{b_1}({1\over{z_1-z_2}}-{1\over{z_1-{\bar{z}}}})^3
({1\over{z_1-z_2}}+{1\over{z_2-{\bar{z}}}})\rbrace}}
Here and everywhere below $z,{\bar{z}}\equiv\pm{i}$.
The next contribution to the amplitude part, quartic in momentum, is given by
\eqn\grav{\eqalign{
<A_2(i)A_0(-i)V_{s=2}(z,{\bar{z}})>_4=
H_{a_1a_2a_3}(p_1)H_{b_1b_2b_3}(p_2)\gamma_{m_1m_2}(p_3)
\cr\times
\int_{-\infty}^\infty
dz_1\int_{-\infty}^\infty{dz_2}(z_1-i)^2(z_2+i)^2{G_{gh}(z_1,z_2,i,-i)}
(-{1\over{z_1-z_2}}+{2\over{z_1-{\bar{z}}}}+{1\over{z_1-z}})
\cr\times
\lbrace{{\eta^{b_3m_2}\eta^{a_3b_2}}
\over{(z_1-z_2)^2(z-z_2)}}
p_2^{a_1}p_2^{a_2}p_1^{b_1}p_1^{m_1}({1\over{z_1-z_2}}-{1\over{z_1-{\bar{z}}}})^2
\cr\times
({1\over{z_1-z_2}}+{1\over{z_2-{\bar{z}}}})
(-{1\over{z_1-{\bar{z}}}}-{1\over{z_2-{\bar{z}}}})
\cr+
{{\eta^{b_3m_2}\eta^{a_3m_1}}
\over{(z_1-{\bar{z}})^2(z-z_2)}}
p_2^{a_1}p_2^{a_2}p_1^{b_1}p_1^{b_2}({1\over{z_1-z_2}}-{1\over{z_1-{\bar{z}}}})^2
({1\over{z_1-z_2}}+{1\over{z_2-{\bar{z}}}})^2
\cr-
{{\eta^{b_3m_2}\eta^{b_2m_1}}
\over{(z_2-{\bar{z}})^2(z-z_2)}}
p_2^{a_1}p_2^{a_2}p_2^{a_3}p_1^{b_1}({1\over{z_1-z_2}}-{1\over{z_1-{\bar{z}}}})^3
({1\over{z_1-z_2}}+{1\over{z_2-{\bar{z}}}})\rbrace}}
The next contribution is
\eqn\grav{\eqalign{
<A_3(i)A_0(-i)V_{s=2}(z,{\bar{z}})>_4=
H_{a_1a_2a_3}(p_1)H_{b_1b_2b_3}(p_2)\gamma_{m_1m_2}(p_3)
\cr\times
\int_{-\infty}^\infty
dz_1\int_{-\infty}^\infty{dz_2}(z_1-i)^2(z_2+i)^2{G_{gh}(z_1,z_2,i,-i)}
\cr\times
\lbrace{{\eta^{b_3m_2}\eta^{a_3b_2}}
\over{(z-z_2)}}
p_2^{a_1}p_2^{a_2}p_1^{b_1}p_1^{m_1}\partial_{z_1}\lbrace{1\over{(z_1-z_2)^2}}
({1\over{z_1-z_2}}-{1\over{z_1-{\bar{z}}}})^2
\cr\times
({1\over{z_1-z_2}}+{1\over{z_2-{\bar{z}}}})
(-{1\over{z_1-{\bar{z}}}}-{1\over{z_2-{\bar{z}}}})\rbrace
\cr+
{{\eta^{b_3m_2}\eta^{a_3m_1}}
\over{(z-z_2)}}
p_2^{a_1}p_2^{a_2}p_1^{b_1}p_1^{b_2}\partial_{z_1}\lbrace
{1\over{(z_1-{\bar{z}})^2}}({1\over{z_1-z_2}}-{1\over{z_1-{\bar{z}}}})^2
({1\over{z_1-z_2}}+{1\over{z_2-{\bar{z}}}})^2\rbrace
\cr-
{{\eta^{b_3m_2}\eta^{b_2m_1}}
\over{(z-z_2)}}
p_2^{a_1}p_2^{a_2}p_2^{a_3}p_1^{b_1}\partial_{z_1}\lbrace
{1\over{(z_2-{\bar{z}})^2}}
({1\over{z_1-z_2}}-{1\over{z_1-{\bar{z}}}})^3
({1\over{z_1-z_2}}+{1\over{z_2-{\bar{z}}}})\rbrace}}
Next,
\eqn\grav{\eqalign{
<A_4(i)A_0(-i)V_{s=2}(z,{\bar{z}})>_4=
2H_{a_1a_2a_3}(p_1)H_{b_1b_2b_3}(p_2)\gamma_{m_1m_2}(p_3)
\cr\times
\int_{-\infty}^\infty
dz_1\int_{-\infty}^\infty{dz_2}(z_1-i)^2(z_2+i)^2{G_{gh}(z_1,z_2,i,-i)}
(-{1\over{z_1-z_2}}+{2\over{z_1-{\bar{z}}}}+{1\over{z_1-z}})
\cr\times\lbrace
(-{{\eta^{a_3b_3}\eta^{a_2m_2}}
\over{(z_1-z_2)(z_1-z)^2}}
+
{{\eta^{a_3m_2}\eta^{a_2b_3}}
\over{(z_1-z)(z_1-z_2)^2}})
p_2^{a_1}p_1^{b_1}p_1^{b_2}p_1^{m_1}
\cr\times
({1\over{z_1-z_2}}-{1\over{z_1-{\bar{z}}}}-{1\over{z_1-z}})
({1\over{z_1-z_2}}+{1\over{z_2-{\bar{z}}}}+{1\over{z_2-z}})^2
({1\over{{z_1-{\bar{z}}}}}+{1\over{{z_2-{\bar{z}}}}})\rbrace}}
Next,

\eqn\grav{\eqalign{
<A_5(i)A_0(-i)V_{s=2}(z,{\bar{z}})>_4=
-H_{a_1a_2a_3}(p_1)H_{b_1b_2b_3}(p_2)\gamma_{m_1m_2}(p_3)
\cr\times
\int_{-\infty}^\infty
dz_1\int_{-\infty}^\infty{dz_2}(z_1-i)^2(z_2+i)^2{G_{gh}(z_1,z_2,i,-i)}
\cr\times\lbrace
(-{{\eta^{a_3b_3}\eta^{a_2m_2}}
\over{(z_1-z_2)(z_1-z)^2}}({1\over{z_1-z_2}}+{2\over{z_1-z}})
\cr
-
{{\eta^{a_3m_2}\eta^{a_2b_3}}
\over{(z_1-z)(z_1-z_2)^2}}({1\over{z_1-{\bar{z}}}}+{2\over{z_1-z_2}}))
p_2^{a_1}p_1^{b_1}p_1^{b_2}p_1^{m_1}
\cr\times
({1\over{z_1-z_2}}-{1\over{z_1-{\bar{z}}}}-{1\over{z_1-z}})
({1\over{z_1-z_2}}+{1\over{z_2-{\bar{z}}}}+{1\over{z_2-z}})^2
({1\over{{z_1-{\bar{z}}}}}+{1\over{{z_2-{\bar{z}}}}})\rbrace}}
Next,
\eqn\grav{\eqalign{
<A_6(i)A_0(-i)V_{s=2}(z,{\bar{z}})>_4=
H_{a_1a_2a_3}(p_1)H_{b_1b_2b_3}(p_2)\gamma_{m_1m_2}(p_3)
\cr\times
\int_{-\infty}^\infty
dz_1\int_{-\infty}^\infty{dz_2}(z_1-i)^2(z_2+i)^2{G_{gh}(z_1,z_2,i,-i)}
(-{1\over{z_1-z_2}}+{2\over{z_1-{\bar{z}}}}+{1\over{z_1-z}})
\cr\times\lbrace
{{({{\eta^{a_3b_3}\eta^{a_4m_2}}}
-
{{\eta^{a_3m_2}\eta^{a_2b_3}}})(p_1)_{a_4}}\over{(z_1-z_2)(z_1-z)}}
\lbrack
{{\eta^{a_2b_2}}\over{(z_1-z_2)^2}}
p_2^{a_1}p_1^{b_1}p_1^{m_1}
\cr\times
({1\over{z_1-z_2}}-{1\over{z_1-{\bar{z}}}}-{1\over{z_1-z}})
({1\over{z_1-z_2}}+{1\over{z_2-{\bar{z}}}}+{1\over{z_2-z}})
({1\over{{z_1-{\bar{z}}}}}+{1\over{{z_2-{\bar{z}}}}})
\cr
+
{{\eta^{a_2m_1}}\over{(z_1-{\bar{z}})^2}}
p_2^{a_1}p_1^{b_1}p_1^{b_2}
\cr\times
({1\over{z_1-z_2}}-{1\over{z_1-{\bar{z}}}}-{1\over{z_1-z}})
({1\over{z_1-z_2}}+{1\over{z_2-{\bar{z}}}}+{1\over{z_2-z}})^2
-
{{\eta^{b_2m_1}}\over{(z_2-{\bar{z}})^2}}
p_2^{a_1}p_2^{a_2}p_1^{b_1}
\cr\times
({1\over{z_1-z_2}}-{1\over{z_1-{\bar{z}}}}-{1\over{z_1-z}})^2
({1\over{z_1-z_2}}+{1\over{z_2-{\bar{z}}}}+{1\over{z_2-z}})
\rbrace}}
Finally,
\eqn\grav{\eqalign{
<A_7(i)A_0(-i)V_{s=2}(z,{\bar{z}})>_4=
H_{a_1a_2a_3}(p_1)H_{b_1b_2b_3}(p_2)\gamma_{m_1m_2}(p_3)
\cr\times
\int_{-\infty}^\infty
dz_1\int_{-\infty}^\infty{dz_2}(z_1-i)^2(z_2+i)^2{G_{gh}(z_1,z_2,i,-i)}
(-{1\over{z_1-z_2}}+{2\over{z_1-{\bar{z}}}}+{1\over{z_1-z}})
\cr\times\lbrace
{{({{{\eta^{a_3b_3}\eta^{a_4m_2}}}\over{(z_1-z_2)(z_1-z)^2}}
-
{{{\eta^{a_3m_2}\eta^{a_2b_3}}}\over{(z_1-z_2)^2(z_1-z)}})(p_1)_{a_4}}}
\cr
\lbrack
{{\eta^{a_2b_2}}\over{(z_1-z_2)^2}}
p_2^{a_1}p_1^{b_1}p_1^{m_1}
\cr\times
({1\over{z_1-z_2}}-{1\over{z_1-{\bar{z}}}}-{1\over{z_1-z}})
({1\over{z_1-z_2}}+{1\over{z_2-{\bar{z}}}}+{1\over{z_2-z}})
({1\over{{z_1-{\bar{z}}}}}+{1\over{{z_2-{\bar{z}}}}})
\cr
+
{{\eta^{a_2m_1}}\over{(z_1-{\bar{z}})^2}}
p_2^{a_1}p_1^{b_1}p_1^{b_2}
\cr\times
({1\over{z_1-z_2}}-{1\over{z_1-{\bar{z}}}}-{1\over{z_1-z}})
({1\over{z_1-z_2}}+{1\over{z_2-{\bar{z}}}}+{1\over{z_2-z}})^2
-
{{\eta^{b_2m_1}}\over{(z_2-{\bar{z}})^2}}
p_2^{a_1}p_2^{a_2}p_1^{b_1}
\cr\times
({1\over{z_1-z_2}}-{1\over{z_1-{\bar{z}}}}-{1\over{z_1-z}})^2
({1\over{z_1-z_2}}+{1\over{z_2-{\bar{z}}}}+{1\over{z_2-z}})
\rbrace}}
This concludes the list of the contributions, quartic in momentum,
to the amplitude (25).
Next, we shall concentrate on the terms, quadratic in $p$.
The straightforward evaluation gives
\eqn\grav{\eqalign{
<A_1(i)A_0(-i)V_{s=2}(z,{\bar{z}})>_2=
-H_{a_1a_2a_3}(p_1)H_{b_1b_2b_3}(p_1)\gamma_{m_1m_2}(p_2)
\cr\times
\int_{-\infty}^\infty
dz_1\int_{-\infty}^\infty{dz_2}(z_1-i)^2(z_2+i)^2{G_{gh}(z_1,z_2,i,-i)}
\cr\times
{{\eta^{a_3b_3}\eta^{a_4m_2}-\eta^{a_4b_3}\eta^{a_4m_2}}\over{(z_1-z)(z_2-{\bar{z}})}}
{\lbrace}
p_2^{a_1}p_2^{a_2}({1\over{z_1-z_2}}-{1\over{z_1-{\bar{z}}}})^2
{{\eta^{a_4b_1}\eta^{b_2m_1}}\over{(z_1-z_2)^2(z_2-{\bar{z}})^2}}
\cr-
p_2^{a_1}p_1^{b_1}
({1\over{z_1-z_2}}-{1\over{z_1-{\bar{z}}}}-{1\over{z_1-z}})
({1\over{z_1-z_2}}+{1\over{z_2-{\bar{z}}}}+{1\over{z_2-z}})
{{\eta^{a_2b_2}\eta^{a_4m_1}}\over{(z_1-z_2)^2(z_2-{\bar{z}})^2}}
\cr+
p_2^{a_1}p_1^{b_1}
({1\over{z_1-z_2}}-{1\over{z_1-{\bar{z}}}}-{1\over{z_1-z}})
(-{1\over{z_1-{\bar{z}}}}+{1\over{z_2-{\bar{z}}}})
{{\eta^{a_2b_2}\eta^{a_4m_1}}\over{(z_1-z_2)^4}}\rbrace}}

The next contribution, quadratic in momentum, is given by

\eqn\grav{\eqalign{
<A_2(i)A_0(-i)V_{s=2}(z,{\bar{z}})>_2=
H_{a_1a_2a_3}(p_1)H_{b_1b_2b_3}(p_2)\gamma_{m_1m_2}(p_3)
\cr\times
\int_{-\infty}^\infty
dz_1\int_{-\infty}^\infty{dz_2}(z_1-i)^2(z_2+i)^2{G_{gh}(z_1,z_2,i,-i)}
\cr\times
{{\eta^{b_3m_2}}\over{(z_2-z)}}
(-{1\over{z_1-z_2}}+{2\over{z_1-{\bar{z}}}}+{1\over{z_1-z}})
\cr\times\lbrace
p_2^{a_1}p_2^{a_2}
{{\eta^{a_3b_1}\eta^{b_2m_1}}\over{(z_1-z_2)^2(z_2-{\bar{z}})^2}}
({1\over{z_1-z_2}}-{1\over{z_1-{\bar{z}}}}-{1\over{z_1-z}})^2
\cr
-
p_2^{a_1}p_1^{b_1}
{{\eta^{a_2b_2}\eta^{a_3m_1}}\over{(z_1-z_2)^2(z_1-{\bar{z}})^2}}
\cr\times
({1\over{z_1-z_2}}-{1\over{z_1-{\bar{z}}}}-{1\over{z_1-z}})
({1\over{z_1-z_2}}+{1\over{z_2-{\bar{z}}}}+{1\over{z_2-z}})
\cr
-
p_2^{a_1}p_1^{m_1}
{{\eta^{a_2b_1}\eta^{a_3b_2}}\over{(z_1-z_2)^4}}
({1\over{z_1-z_2}}-{1\over{z_1-{\bar{z}}}}-{1\over{z_1-z}})
\cr\times
({1\over{z_1-z_2}}+{1\over{z_2-{\bar{z}}}}+{1\over{z_2-z}})
(-{1\over{z_1-{\bar{z}}}}+{1\over{z_2-{\bar{z}}}})\rbrace}}
Next,
\eqn\grav{\eqalign{
<A_3(i)A_0(-i)V_{s=2}(z,{\bar{z}})>_2=
H_{a_1a_2a_3}(p_1)H_{b_1b_2b_3}(p_2)\gamma_{m_1m_2}(p_3)
\cr\times
\int_{-\infty}^\infty
dz_1\int_{-\infty}^\infty{dz_2}(z_1-i)^2(z_2+i)^2{G_{gh}(z_1,z_2,i,-i)}
\cr\times
{{\eta^{b_3m_2}}\over{(z_2-z)}}
\cr\times\lbrace
p_2^{a_1}p_2^{a_2}\partial_{w_1}\lbrace
{{\eta^{a_3b_1}\eta^{b_2m_1}}\over{(z_1-z_2)^2(z_2-{\bar{z}})^2}}
({1\over{z_1-z_2}}-{1\over{z_1-{\bar{z}}}}-{1\over{z_1-z}})^2
\rbrace
\cr
-
p_2^{a_1}p_1^{b_1}
\partial_{w_1}\lbrace
{{\eta^{a_2b_2}\eta^{a_3m_1}}\over{(z_1-z_2)^2(z_1-{\bar{z}})^2}}
({1\over{z_1-z_2}}-{1\over{z_1-{\bar{z}}}}-{1\over{z_1-z}})
\cr\times
({1\over{z_1-z_2}}+{1\over{z_2-{\bar{z}}}}+{1\over{z_2-z}})
\rbrace
\cr
-
p_2^{a_1}p_1^{m_1}
\partial_{w_1}\lbrace
{{\eta^{a_2b_1}\eta^{a_3b_2}}\over{(z_1-z_2)^4}}
({1\over{z_1-z_2}}-{1\over{z_1-{\bar{z}}}}-{1\over{z_1-z}})
\cr\times
({1\over{z_1-z_2}}+{1\over{z_2-{\bar{z}}}}+{1\over{z_2-z}})
(-{1\over{z_1-{\bar{z}}}}+{1\over{z_2-{\bar{z}}}})\rbrace\rbrace}}
Next,
\eqn\grav{\eqalign{
<A_4(z_1)A_0(z_2)V_{s=2}(z,{\bar{z}})>_2=
-2H_{a_1a_2a_3}(p_1)H_{b_1b_2b_3}\gamma_{m_1m_2}
\cr\times
\int_{-\infty}^\infty
dz_1\int_{-\infty}^\infty{dz_2}(z_1-i)^2(z_2+i)^2{G_{gh}(z_1,z_2,i,-i)}
(-{1\over{z_1-z_2}}+{2\over{z_1-{\bar{z}}}}+{1\over{z_1-z}})
\cr\times
(-{{\eta^{a_3b_3}\eta^{a_2m_2}}\over{(z_1-z_2)(z_1-{{z}})^2}}
+
{{\eta^{a_2b_2}\eta^{a_3m_2}}\over{(z_1-z_2)^2(z_1-{{z}})}})
\cr\times\lbrace
-p_1^{b_1}p_2^{a_1}({1\over{z_1-z_2}}-{1\over{z_1-{\bar{z}}}}-{1\over{z_1-z}})
({1\over{z_1-z_2}}+{1\over{z_2-{\bar{z}}}}+{1\over{z_2-z}})
{{\eta^{b_2m_1}}\over{(z_2-{\bar{z}})^2}}
\cr+
-p_1^{b_1}p_1^{b_2}
({1\over{z_1-z_2}}+{1\over{z_2-{\bar{z}}}}+{1\over{z_2-z}})^2
{{\eta^{a_1m_1}}\over{(z_1-{\bar{z}})^2}}
\cr
-p_1^{b_1}p_1^{m_1}({1\over{z_2-{\bar{z}}}}-{1\over{z_1-{\bar{z}}}})
({1\over{z_1-z_2}}+{1\over{z_2-{\bar{z}}}}+{1\over{z_2-z}})
{{\eta^{a_1b_2}}\over{(z_1-z_2)^2}}
\rbrace}}
Next,
\eqn\grav{\eqalign{
<A_5(i)A_0(-i)V_{s=2}(z,{\bar{z}})>_2=
-2H_{a_1a_2a_3}(p_1)H_{b_1b_2b_3}(p_2)\gamma_{m_1m_2}(p_3)
\cr\times
\int_{-\infty}^\infty
dz_1\int_{-\infty}^\infty{dz_2}(z_1-i)^2(z_2+i)^2{G_{gh}(z_1,z_2,i,-i)}
\cr\times
({{\eta^{a_3b_3}\eta^{a_2m_2}}\over{(z_1-z_2)(z_1-{{z}})^2}}
({1\over{z_1-z_2}}+{2\over{w_2-z}})
-
{{\eta^{a_2b_2}\eta^{a_3m_2}}\over{(z_1-z_2)^2(z_1-{{z}})}})
({2\over{z_1-z_2}}+{1\over{w_2-z}})
\cr\times\lbrace
-p_1^{b_1}p_2^{a_1}({1\over{z_1-z_2}}-{1\over{z_1-{\bar{z}}}}-{1\over{z_1-z}})
({1\over{z_1-z_2}}+{1\over{z_2-{\bar{z}}}}+{1\over{z_2-z}})
{{\eta^{b_2m_1}}\over{(z_2-{\bar{z}})^2}}
\cr+
-p_1^{b_1}p_1^{b_2}
({1\over{z_1-z_2}}+{1\over{z_2-{\bar{z}}}}+{1\over{z_2-z}})^2
{{\eta^{a_1m_1}}\over{(z_1-{\bar{z}})^2}}
\cr
-p_1^{b_1}p_1^{m_1}({1\over{z_2-{\bar{z}}}}-{1\over{z_1-{\bar{z}}}})
({1\over{z_1-z_2}}+{1\over{z_2-{\bar{z}}}}+{1\over{z_2-z}})
{{\eta^{a_1b_2}}\over{(z_1-z_2)^2}}
\rbrace}}
Next,
\eqn\grav{\eqalign{
<A_6(i)A_0(-i)V_{s=2}(z,{\bar{z}})>_2=
H_{a_1a_2a_3}(p_1)H_{b_1b_2b_3}(p_2)\gamma_{m_1m_2}(p_3)
\cr\times
\int_{-\infty}^\infty
dz_1\int_{-\infty}^\infty{dz_2}(z_1-i)^2(z_2+i)^2{G_{gh}(z_1,z_2,i,-i)}
(-{1\over{z_1-z_2}}+{2\over{z_1-{\bar{z}}}}+{1\over{z_1-z}})
\cr\times
{{(p_1)_{\alpha_4}}\over{(z_1-z_2)(z_1-z)}}(-{{\eta^{a_3b_3}\eta^{a_4m_2}}}
+\eta^{a_4b_3}\eta^{a_3m_2})
\cr\times\lbrace
{{p_2^{a_1}\eta^{a_2b_1}\eta^{b_2m_1}}\over{(z_1-z_2)^2(z_2-{\bar{z}})^2}}
({1\over{z_1-z_2}}-{1\over{z_1-{\bar{z}}}}-{1\over{z_1-z}})
\cr
-
{{p_1^{b_1}\eta^{a_1b_2}\eta^{a_2m_1}}\over{(z_1-z_2)^2(z_1-{\bar{z}})^2}}
({1\over{z_1-z_2}}+{1\over{z_2-{\bar{z}}}}+{1\over{z_2-z}})
\cr
+
{{p_1^{m_1}\eta^{a_1b_1}\eta^{a_2b_2}}\over{(z_1-z_2)^4}}
({1\over{z_2-{\bar{z}}}}+{1\over{z_1-{\bar{z}}}})
\rbrace
}}
Finally,
\eqn\grav{\eqalign{
<A_7(i)A_0(-i)V_{s=2}(z,{\bar{z}})>_2=
H_{a_1a_2a_3}(p_1)H_{b_1b_2b_3}(p_2)\gamma_{m_1m_2}(p_3)
\cr\times
\int_{-\infty}^\infty
dz_1\int_{-\infty}^\infty{dz_2}(z_1-i)^2(z_2+i)^2{G_{gh}(z_1,z_2,i,-i)}
\cr\times
{{(p_1)_{\alpha_4}}\over{(z_1-z_2)(z_1-z)}}
(-{{\eta^{a_3b_3}\eta^{a_4m_2}}\over{z_1-z}}
+{{\eta^{a_4b_3}\eta^{a_3m_2}}\over{z_1-z_2}})
\cr\times\lbrace
{{p_2^{a_1}\eta^{a_2b_1}\eta^{b_2m_1}}\over{(z_1-z_2)^2(z_2-{\bar{z}})^2}}
({1\over{z_1-z_2}}-{1\over{z_1-{\bar{z}}}}-{1\over{z_1-z}})
\cr
-
{{p_1^{b_1}\eta^{a_1b_2}\eta^{a_2m_1}}\over{(z_1-z_2)^2(z_1-{\bar{z}})^2}}
({1\over{z_1-z_2}}+{1\over{z_2-{\bar{z}}}}+{1\over{z_2-z}})
\cr
+
{{p_1^{m_1}\eta^{a_1b_1}\eta^{a_2b_2}}\over{(z_1-z_2)^4}}
({1\over{z_2-{\bar{z}}}}+{1\over{z_1-{\bar{z}}}})\rbrace}}
This concludes the list of terms,quadratic in momentum.
Finally, we shall list the contributions of the order zero in
momentum.
We obtain
\eqn\grav{\eqalign{
<A_1(i)A_0(-i)V_{s=2}(z,{\bar{z}})>_0=
-H_{a_1a_2a_3}(p_1)H_{b_1b_2b_3}(p_2)\gamma_{m_1m_2}(p_3)
\cr\times
\int_{-\infty}^\infty
dz_1\int_{-\infty}^\infty{dz_2}(z_1-i)^2(z_2+i)^2{G_{gh}(z_1,z_2,i,-i)}
\cr\times\lbrace
{{(\eta^{a_4b_3}\eta^{a_3m_2}-\eta^{a_3b_3}\eta^{a_4m_2})
\eta_{a_4}^{b_2}\eta^{a_1b_1}\eta^{a_2m_1}}\over
{(z_1-z_2)^5(z_1-{\bar{z}})^2(z_1-z)}}\rbrace}}
Then,

\eqn\grav{\eqalign{
<A_2(i)A_0(-i)V_{s=2}(z,{\bar{z}})>_0=
-H_{a_1a_2a_3}(p_1)H_{b_1b_2b_3}(p_2)\gamma_{m_1m_2}(p_3)
\cr\times
\int_{-\infty}^\infty
dz_1\int_{-\infty}^\infty{dz_2}(z_1-i)^2(z_2+i)^2{G_{gh}(z_1,z_2,i,-i)}
(-{1\over{z_1-z_2}}+{2\over{z_1-{\bar{z}}}}+{1\over{z_1-z}})
\cr\times
{{\eta^{a_1b_1}\eta^{a_2b_2}\eta^{a_3m_1}\eta^{b_3m_2}}\over{
{(z_1-z_2)^4(z_1-{\bar{z}})^2(z_1-z)}}}}}
Next,
\eqn\grav{\eqalign{
<A_3(i)A_0(-i)V_{s=2}(z,{\bar{z}})>_0=
-H_{a_1a_2a_3}(p_1)H_{b_1b_2b_3}(p_2)\gamma_{m_1m_2}(p_3)
\cr\times
\int_{-\infty}^\infty
dz_1\int_{-\infty}^\infty{dz_2}(z_1-i)^2(z_2+i)^2{G_{gh}(z_1,z_2,i,-i)}
\cr\times\partial_{w_1}\lbrace
{{\eta^{a_1b_1}\eta^{a_2b_2}\eta^{a_3m_1}\eta^{b_3m_2}}\over{
{(z_1-z_2)^4(z_1-{\bar{z}})^2(z_1-z)}}}\rbrace}}
Next,
\eqn\grav{\eqalign{
<A_4(i)A_0(-i)V_{s=2}(z,{\bar{z}})>_0=
-4H_{a_1a_2a_3}(p_1)H_{b_1b_2b_3}(p_2)\gamma_{m_1m_2}(p_3)
\cr\times
\int_{-\infty}^\infty
dz_1\int_{-\infty}^\infty{dz_2}(z_1-i)^2(z_2+i)^2{G_{gh}(z_1,z_2,i,-i)}
(-{1\over{z_1-z_2}}+{2\over{z_1-{\bar{z}}}}+{1\over{z_1-z}})
\cr\times
{{\eta^{a_1b_1}\eta^{b_2m_2}}\over{(z_1-z_2)^2(z_2-z)^2}}
({{{\eta^{a_2m_1}}\eta^{a_3b_3}}\over{(z_1-{\bar{z}})^2(z_1-z_2)}}
-
{{{\eta^{a_2b_3}}\eta^{a_3m_1}}\over{(z_1-{\bar{z}})(z_1-z_2)^2}})}}
Finally,
\eqn\grav{\eqalign{
<A_5(i)A_0(-i)V_{s=2}(z,{\bar{z}})>_0=
-2H_{a_1a_2a_3}(p_1)H_{b_1b_2b_3}(p_2)\gamma_{m_1m_2}(p_3)
\cr\times
\int_{-\infty}^\infty
dz_1\int_{-\infty}^\infty{dz_2}(z_1-i)^2(z_2+i)^2{G_{gh}(z_1,z_2,i,-i)}
\cr\times\partial_{w_1}\lbrace
{{\eta^{a_1b_1}\eta^{b_2m_2}}\over{(z_1-z_2)^2(z_2-z)^2}}
({{{\eta^{a_2m_1}}\eta^{a_3b_3}}\over{(z_1-{\bar{z}})^2(z_1-z_2)}}
-
{{{\eta^{a_2b_3}}\eta^{a_3m_1}}\over{(z_1-{\bar{z}})(z_1-z_2)^2}})\rbrace
\cr
<A_6(i)A_0(-i)V_{s=2}(z,{\bar{z}})>_0=<A_7(i)A_0(-i)V_{s=2}(z,{\bar{z}})>_0=0}}

This concludes the list of all the contributions
 to the spin 3-graviton scattering, up to
the terms, quartic in momentum.
The next step is to evaluate the $z_1,z_2$-integrals , giving the
relative coefficients in front of all the terms, listed in (27)-(45).
To perform the integration, it is convenient to
introduce the regulator $e^{i\lambda(z_1+z_2)}(\lambda>0)$ inside
each of the integrals (setting $\lambda\rightarrow{0}$ upon
the calculation).
 The regulator ensures that the contour integrals
over cemicircle of radius $R\rightarrow\infty$ in the upper half-plane vanish
and therefore the straight line integrals can be obtained from the
the appropriate residues in the $z_1,z_2$-integrals.
Shifting the straight line  $z_1,z_2$ contours according to
 $z_1\rightarrow{z_1-i0},z_2\rightarrow{z_2+i0}$, it is convenient
to evaluate the integral in $z_2$ first and then in $z_1$.
The evaluation of the $z_1,z_2$ integrals is then straightforward;
 transforming back
to the position space we find that the total 
expression for the three-vertex of two  $s=3$ particles and a graviton
from the three-point amplitude on the disc is given by:
\eqn\grav{\eqalign{-4\pi^2A_{s=3,3,2}=
-2\eta^{a_3b_3}\eta^{a_1m_1}\partial_{b_1}\partial_{b_2}
\partial_{m_2}H^{a_1a_2a_3}\partial_{a_2}H^{b_1b_2b_3}\gamma^{m_1m_2}
\cr
+2\eta_{a_1m_1}\eta_{a_2m_2}\eta^{a_4b_4}
\partial_{b_1}\partial_{b_2}\partial_{b_4}H^{a_1a_2a_3}\partial_{a_4}
H^{b_1b_2b_3}\gamma^{m_1m_2}
\cr
+2\eta^{a_1m_1}\eta^{a_2m_2}
\partial_{b_1}\partial_{b_2}\partial_{b_3}H^{a_1a_2a_3}\partial_{a_3}
H^{b_1b_2b_3}\gamma^{m_1m_2}
\cr
+
\eta_{a_2b_2}\eta_{a_3b_3}
\partial_{b_1}\partial_{m_1}\partial_{m_2}H^{a_1a_2a_3}\partial_{a_1}
H^{b_1b_2b_3}\gamma^{m_1m_2}
\cr
-2\eta_{a_1m_1}\eta_{a_2b_3}
\partial_{b_1}\partial_{b_2}\partial_{m_2}H^{a_1a_2a_3}\partial_{a_3}
H^{b_1b_2b_3}\gamma^{m_1m_2}
\cr
-2\eta_{a_2m_1}\eta_{b_2m_2}
\partial_{b_1}\partial_{b_3}H^{a_1a_2a_3}\partial_{a_1}\partial_{a_3}
H^{b_1b_2b_3}\gamma^{m_1m_2}
\cr
-2\eta_{a_2m_1}\eta_{b_2m_2}\eta_{a_1b_3}\eta^{a_4b_4}
\partial_{b_1}\partial_{a_4}H^{a_1a_2a_3}\partial_{a_3}\partial_{b_4}
H^{b_1b_2b_3}\gamma^{m_1m_2}
\cr
+
\eta_{a_2b_2}\eta_{a_3b_3}\eta_{b_1m_1}\eta^{a_4b_4}
\partial_{m_2}\partial_{a_4}H^{a_1a_2a_3}\partial_{a_1}\partial_{b_4}
H^{b_1b_2b_3}\gamma^{m_1m_2}
\cr
+2
\eta_{a_1m_1}\eta_{a_2m_2}\eta_{a_3b_2}\eta^{a_4b_4}
\partial_{b_1}\partial_{b_3}\partial_{a_4}H^{a_1a_2a_3}\partial_{b_4}
H^{b_1b_2b_3}\gamma^{m_1m_2}
\cr
-2\eta_{a_1m_1}\eta_{a_2m_2}\eta_{a_3b_1}\eta^{a_4b_4}
\partial_{b_2}\partial_{a_4}H^{a_1a_2a_3}\partial_{a_2}\partial_{b_4}
H^{b_1b_2b_3}\gamma^{m_1m_2}
\cr
+
\eta_{a_1m_2}\eta_{a_2b_2}\eta_{a_3b_3}\eta^{a_4b_4}
\partial_{b_1}\partial_{m_1}\partial_{a_4}H^{a_1a_2a_3}\partial_{b_4}
H^{b_1b_2b_3}\gamma^{m_1m_2}
\cr
-\eta_{a_1m_1}\eta_{b_3m_2}
\partial_{b_1}\partial_{b_2}H^{a_1a_2a_3}\partial_{a_2}
\partial_{a_3}H^{b_1b_2b_3}\gamma^{m_1m_2}
\cr
-
2\eta_{b_1m_1}\eta_{a_2m_2}\eta_{a_3b_2}\eta^{a_4b_4}
\partial_{b_3}\partial_{a_4}H^{a_1a_2a_3}\partial_{a_1}\partial_{b_4}
H^{b_1b_2b_3}\gamma^{m_1m_2}
\cr
+
2\eta_{a_1m_1}\eta_{a_2m_2}\eta_{a_3b_3}\eta^{a_4b_4}
\partial_{b_1}\partial_{b_2}\partial_{a_4}H^{a_1a_2a_3}\partial_{b_4}
H^{b_1b_2b_3}\gamma^{m_1m_2}
\cr
+
2\eta_{a_2m_1}\eta_{a_3m_2}
\partial_{b_1}\partial_{b_2}\partial_{b_3}H^{a_1a_2a_3}\partial_{a_1}
H^{b_1b_2b_3}\gamma^{m_1m_2}
\cr
+
2\eta_{a_1m_1}\eta_{b_1m_2}\eta_{a_2b_2}\eta_{a_3b_3}\eta^{a_4b_4}\eta^{a_5b_5}
\partial_{a_4}\partial_{a_5}H^{a_1a_2a_3}\partial_{b_4}
\partial_{b_5}H^{b_1b_2b_3}\gamma^{m_1m_2}}}
Contracting the indices and using the
conditions (2), it is straightforward to show that, modulo
partial integration 
one can cast the expression (46)  as
\eqn\grav{\eqalign{
{-4\pi^2A_{s=3,3,2}}
\cr
=w_{abcd}\lbrace
2h^a_{mn}\partial^d\partial^n{h}^{bcm}
-2\partial^m{h^{acn}}\partial_mh^{bd}_n-2\partial^n{h^{acm}}
\partial_mh^{bd}-h^{a}_{mn}\partial^b\partial^dh^{cmn}\rbrace
}}
where $w_{abcd}(\gamma)$ is linearized Weyl tensor.
Remarkably, the terms (34)-(40), quadratic in derivatives, as well as those
of (41)-(45) containing no derivatives, drop out as a result of the integration,
so only the terms, quartic in derivatives remain. 
It is instructive to compare the expressions (46), (47) to the interaction
vertex given in the important
papers ~{\nba, \nbb} (this vertex is also a flat limit of the one
found in ~{\fradkin} for the AdS case)
 According 
to {\nba, \nbb}, the interaction
of graviton with massless spin 3 fields is determined by the linearized
Weyl tensor times the quadratic combination of spin 3 fields. The 
3-vertex (46), (47) following from the string scattering amplitude on the disc,
reproduces
 the 3-vertex found in ~{\nba, \nbb} with the gauge partially
fixed according to the constraints (2).
 The fact that string theory gives us the gauge fixed version
of the cubic vertex of the spin 3 - graviton interaction, is not surprising, 
if we compare the higher spin case to the standard example of a photon when
the BRST (transversality) constraints on the vertex operator 
naturally imply the Lorenz
gauge choice for the space-time field.

\centerline {\bf 4. Conclusion and Discussion}

In this paper we have  determined the gravitational
coupling of the massless spin 3 field from string 
theory, by computing the appropriate 
correlation function  on the disc. Remarkably, the string theory 
calculation turns out to  reproduce (up to partial gauge fixing) the 
non-Abelian 
cubic coupling of the spin 3 field with the graviton through the linearized 
Weyl tensor, derived in ~{\nba, \nbb}, which also is the flat limit of the
Fradkin-Vasiliev vertex in the frame-like approach ~{\fradkin}
This result is in agreement with more general property of spin $s$
fields coupling to the gravity, with the non-Abelian 
cubic interaction vertices containing
 not less than $2s-2$ space-time derivatives.
Note that, although the lower derivative terms initially appear
in the calculation, they all drop out as a result of the integration
(recall that the positive picture expressions for the higher spin operators
exist in the integrated form only)
This altogether is an important consistency test for the string vertex 
operators for the higher spins, considered in this work.
It would be important to generalize this disc calculation to particles
with the spins $s\geq{4}$ interacting with the graviton, in order
to check the $2s-2$ derivative rule for $s>3$.
Generally, one would expect each of the correlators
to produce three principal contributions, differing in the number
of the space-time derivatives ($2s-2,2s$ and $2s+2$), according to
corresponding to three possible cubic couplings 
of spin $s$ particles and the graviton ~{\metsaev}.
Although we concentrated on the $3$-point amplitude
 of two spin $3$ particles with the graviton on the disc,
the amplitude, considered in this paper is structurally similar
to the 4-point amplitude in open string theory, involving
two photons and 2 spin 3 particles. Generally, the quartic terms
in the interacting higher spin field theories are cumbersome
and their structure is not yet well understood. It would be
important to investigate the quartic couplings by using the vertex
operator computations in string theory, which appears to be an efficient
and a promising framework to approach the problem.

\centerline{\bf Acknowledgements}

I would like express my gratitude to Augusto Sagnotti and
other members of the group at SNS in Pisa for the hospitality
and the productive discussions
during my visit to Scuola Normale in January 2010, where the initial
stages of this work were completed. 
It is a pleasure to thank Nick Boulanger and Per Sundell for illuminating
and productive discussions on the vertex operators 
and the non-Abelian couplings for the higher spin fields.
In particular, I would like to thank Per Sundell for pointing out 
to me a number of interesting problems related to non-Abelian couplings
of the higher spin fields.
I also would like to thank Yoshihisa Kitazawa for the hospitality during
my visit to KEK in February 2010, where parts of the calculation,
presented in this work, have been done.

\listrefs

\end